\documentclass[]{elsarticle}
\usepackage[dvipsnames]{xcolor}

\usepackage[utf8]{inputenc}
\usepackage{tcolorbox}
\newtcbox{\inlinecode}{on line, boxrule=0pt, boxsep=0pt, top=2pt, left=2pt, bottom=2pt, right=2pt, colback=gray!15, colframe=white, fontupper={\ttfamily \footnotesize}}
\newtcbox{\inlinebash}{on line, boxrule=0pt, boxsep=0pt, top=2pt, left=2pt, bottom=2pt, right=2pt, colback=gray!15, colframe=white, fontupper={\ttfamily \footnotesize}}
\usepackage{paralist}
\usepackage{tikz}

\usepackage[draft,author=,mode=multiuser]{fixme}
\fxsetup{theme=colorsig}
\fxsetface{margin}{\tiny}
\fxsetface{env}{\footnotesize}
\fxsetface{inline}{\footnotesize}
\FXRegisterAuthor{p}{mp}{pizzo}
\FXRegisterAuthor{d}{dp}{Diego}

\usepackage{array}					\usepackage{siunitx}	\usepackage{listings}	
\usepackage{collcell}

\usepackage{colortbl}
\usepackage{algpseudocode}
\usepackage{algorithm}
\algnewcommand{\LineComment}[1]{\State \(\triangleright\) #1}
\algnewcommand{\StateBash}[1]{\State \inlinebash{#1}}

\newcolumntype{I}{>{\centering\arraybackslash\collectcell\inlinecode}c<{\endcollectcell}}
\newcolumntype{M}[1]{>{\centering\arraybackslash}m{#1}}

\usepackage{hyperref}
\usepackage{textcomp}
\usepackage[inline,shortlabels]{enumitem}
\usepackage{adjustbox}
\usepackage{multirow}
\usepackage{makecell}
\usepackage{mathtools}

\DeclarePairedDelimiter\floor{\lfloor}{\rfloor}

\journal{Blockchain: Research and Applications}

\begin{document}

	\begin{frontmatter}

	\title{Toward Scalable Docker-Based Emulations of Blockchain Networks for Research and Development\tnoteref{t1}}

	\tnotetext[t1]{A preliminary version of this paper was published in the proceeding of the ``DLT’23: 5th Distributed Ledger Technology Workshop'' under the title ``Toward Scalable Docker-Based Emulations of Blockchain Networks''~\cite{Pennino2023ScalableEmulations}.}

	\author[inst1,inst2]{Diego Pennino\corref{cor1}}
	\ead{diego.pennino@unitus.it}
	
	\cortext[cor1]{Corresponding author}

	\author[inst2]{Maurizio Pizzonia} 	\ead{pizzonia@ing.uniroma3.it}

	\affiliation[inst1]{
		organization={Università degli Studi della Tuscia, Dipartimento di Economia, Ingegneria, Società e Impresa},
		addressline={Via del paradiso 47},
		postcode={01100},
		city={Viterbo},
		country={Italy}}
	
	\affiliation[inst2]{
		organization={Università degli Studi Roma Tre, Dipartimento di Ingegneria Civile, Informatica e delle Tecnologie Aeronautiche},
		addressline={Via della Vasca Navale 79},
		postcode={00146},
		city={Rome},
		country={Italy}}

	\begin{abstract}
	Blockchain, like any other complex technology, needs a strong testing methodology to
	support its evolution in both research and development contexts. Setting up meaningful 
	tests for permissionless blockchain technology is a notoriously
	complex task for several reasons: software is complex, large number of nodes
	are involved, network is non ideal, etc. Developers usually adopt small
	virtual laboratories or costly real devnets, based on real software.
	Researchers usually prefer simulations of a large number of nodes, based on
	simplified models.
	
	In this paper, we aim to obtain the advantages of both approaches, i.e.,
	performing large, realistic, inexpensive, and flexible experiments, using real
	blockchain software within a virtual environment. To do that, we tackle the
	challenge of running large blockchain networks in a single physical machine,
	leveraging Linux and Docker. We analyze a number of problems that arise when
	large blockchain networks are emulated and we provide technical solutions for
	all of them. Finally, we describe two experiences of emulating fairly large
	blockchain networks on a single machine, adopting both research oriented and
	production oriented software, and involving up to more than 3000 containers.
\end{abstract}


	\end{frontmatter}

	\section{Introduction}

Performing realistic experiments for blockchain networks is notoriously hard.
However, reproducing a realistic blockchain network is desirable for both
research and development purposes. Researchers may aim to test new protocols and
make realistic measurements in laboratory. Developers of blockchain software
would like to test new software versions before distribution and deployment and
possibly perform what-if analysis without disrupting costly devnets. 

The complexity of setting up realistic experiments stems from several factors.

\begin{enumerate}

\item \label{f:large} A real blockchain network may encompass a very large number of nodes.

\item \label{f:complex} The software run by nodes is usually quite complex.

\item \label{f:transport} Communications among nodes are affected by typical
properties of transport protocols (e.g., the \emph{slow start} of TCP).

\item \label{f:delay} Nodes are usually spread over the internet, which
implies that any communication between them is affected by all wanted and
unwanted properties of the real internet, prominently delay and packet loss.

\end{enumerate}

Item~\ref{f:large}, naturally leads us toward the adoption of a simplified and clean simulation. On the
contrary, Items~\ref{f:complex}, \ref{f:transport}, and~\ref{f:delay} may be
better reproduced by emulation environments that leverage the very same
technology and software of a real production environment.

Currently, developers adopt development networks that mimic to some extent
production environments in the sense that they are made of a substantial number
of real nodes, are spread over the internet, and run the real software. However,
this approach is costly (because many machines are dedicated to this task), time-consuming
 (since machines have to be managed), and unhandy (because when a new
version has to be tested all machines have to be updated making it potentially unusable for other purposes). Further, sharing the network among tests of several software
versions can lead to results that are hard to interpret. On the other hand,
researchers mostly limit themselves to simplified simulations to reduce costs.

In this paper, we show how it is possible to run thousands of distinct
blockchain nodes on a single machine, running realistic software and
adopting a real TCP/IP stack. 
We limit the perimeter of our investigation to running the emulation on a single machine, leaving 
the investigation about the adoption of more than one machine to further research work.
We also show how to inject realistic delays into the emulated
network to also emulate realistic timings.

Our approach is simple in principle but not so easy to be applied in practice.
We just create one Docker~\cite{docker} container for each node, each with its
own IP address and let them talk to each other.
Practically applying this natural approach has a number of complex aspects. 

The main contribution of our work is \begin{inparaenum}[(1)]
\item a list of technical problems we encountered
that limit the number of nodes we can emulate and \item corresponding recipes we suggest to solve them.
\end{inparaenum}
 In particular, we analyze and address the
limits of the Linux kernel related to launching and connecting a large number of
containers. We deal with the detrimental effect of the ARP protocol realizing a
solution that completely removes this kind of traffic without resorting to a
quadratic number of static ARP cache entries. We show how to configure
realistic, internet-like, delays among all pairs of nodes without inserting a
quadratic number of firewalling rules. We show how it is possible to arbitrarily
reduce CPU consumption of our experimentation by inflating all the time-related
parameters involved in the emulation. This is not straightforward, since slowing
down the network and the software may trigger unexpected timeouts. In
particular, we show how to inflate TCP retransmission timeouts, which are hard-coded
in the kernel. Since recompiling the kernel is cumbersome and time consuming, we
show a technique to change it without resorting to kernel recompilation.

Applying all the above mentioned approaches, we show that it is possible to run
very large blockchain emulations on a single machine with typical production
software or with research-oriented software. In particular, we aim to show that,
with typical hardware and software, the scalability bottleneck is the RAM, since
all other limits, comprising CPU, can be addressed programmatically.

We describe two large emulation experiments whose intent is to apply the described techniques 
in practice and show that the limiting resource is actual the amount of available RAM.
The first experiment consists of the
emulation of a research-targeted blockchain realized in Python consisting of
more than 3000 containers occupying about 350 GiB or RAM. The second one consists
of the emulation of a PoS-based Ethereum 2 network running plain versions of the
Go-Ethereum~\cite{geth} and Prysm~\cite{prysm} software in a typical
configuration. This second experiment comprises 750 nodes occupying about 307 GiB
of RAM. For this emulation, we provide a companion GitHub repository~\cite{companion-repo}
with scripts and instructions to reproduce the experiment.

The rest of this paper is structured as follows.
In Section~\ref{sec:SoA}, we review the state of the art.
In Section~\ref{sec:context}, we describe the two experimentation contexts for which we have undertaken our work.
In Section~\ref{sec:launching}, we show how to launch a large number of blockchain nodes on a single machine.
In Section~\ref{sec:connecting}, we show how to connect them.
In Section~\ref{sec:autoarpd}, we describe the ARP traffic problem and our AutoARPD tool to solve it. 
In Section~\ref{sec:latency}, we describe the configurations to emulate internet-like delays among nodes.
In Section~\ref{sec:inflation}, we describe how to arbitrarily lower CPU load by time inflation.
In Section~\ref{sec:experiments}, we describe in detail our two emulation experiences discussing the practical application of the techniques described in this paper.
In Section~\ref{sec:conclusions}, we draw the conclusions.

	\section{State of the Art}\label{sec:SoA}

As stated in the introduction, currently blockchain developers adopt real
networks, usually called \emph{devnets}, dedicated to experiment with new
releases. The drawbacks of this approach were listed in the introduction. Before
deployment on a devnet, it is likely that developers perform some small scale
test of the software in a small laboratory environment, possibly using some form
of virtualization. In this paper, we essentially discuss how to scale this last
approach to a large number of nodes. Researchers mostly use \emph{simulation},
that is ad-hoc software based on simplified models of the elements of the
blockchain system that computes its ``evolution'' over time. 
A large number of simulation systems are described in literature or
freely available for download over the internet, see for example, ~\cite{dlsf,
agrawal2020blocksim, evibesPlasma, fattahi2020simba, aoki2019simblock,
lathif2018cidds, hive}. The work in~\cite{dinh2017blockbench} provides a
framework to evaluate private blockchains technologies based on six-layers:
application, contract, incentive, consensus, node/data, network. One of the most
relevant and accurate emulation/simulation approach is called \emph{BlockPerf}
and it is proposed in~\cite{polge2021blockperf}. It emulates the network layer
by taking advantage of geographically sparse real nodes and simulates the
remaining layers, trying to cover as much as possible the layers mentioned
above. This approach has the disadvantage of relying on many geographically
distributed machines which is one of the problems of the devnet approach. 
Further, it requires careful planning and deployment of nodes. This problem is
even more relevant for IoT-targeted technologies, like the
\emph{tangle}~\cite{popov2018tangle,Tangle2.0} in IOTA, where emulation of a
very large number of devices might be required for realistic experimentation.

The difficulty of blockchain emulation is also remarked in~\cite{s20123358}.
The authors of that survey, point out that this kind of emulation is extremely
demanding in terms of resources and that no general tool exists to support it.
In our work, we aim to provide approaches for general blockchain emulations
that help to overcome these difficulties.

Regarding peer-to-peer networks, mostly targeted to file/content storage and
distribution, many simulation approaches were proposed (several are reported
in~\cite{basu2013state,ebrahim2014peer}). On the contrary, proposals based on the emulation
approach are rare~\cite{nussbaum2006lightweight}.

Further, the work in~\cite{5280241} proposes a model for generating large
realistic internet delay matrices with the intent to support the simulation of large
geographically distributed systems. We use the results of that work in
Section~\ref{sec:latency}.

	\section{Our Two Experimental Contexts}\label{sec:context}

As mentioned in the introduction, the aims of our experimentation are twofold: 
\begin{inparaenum}[(1)]
\item shows a practical application of the techniques described in the paper and \item shows that it is possible to scale emulation so that the remaining bottleneck is provided by the available amount of RAM.
\end{inparaenum}
In this section, we describe two experimental contexts in which large scale
blockchain emulation is desirable. 

The first context is related to research in blockchain scalability and in
particular in storing the blockchain data in a Distribute Hash Table (DHT). In
this experiment, the techniques to test are too complex for the simulation
approach to be viable. On the other hand, scaling with respect to the number of
nodes is one of the objectives of the research. In this context, the software is
lightweight and under the control of the researcher.

The second context is related to the testing of a realistic permissionless
blockchain network (a PoS-based Ethereum 2 network) implemented by running
standard production-ready software. We show that running a third-party
production-ready software introduces a number of tricky aspects.

In this section, we just describe the experimental contexts, the aims, and the
foreseen problems to motivate the study of the techniques introduced in the rest
of the paper. We describe the actual results of our experiments for these two
contexts in Section~\ref{sec:experiments}.

\subsection{Research in Blockchain Scalability}\label{sec:context-scientific}

In our first context, we aim to support experimental research regarding 
new storage approaches to be adopted in new permissionless blockchain systems. It is well known that full
nodes in a blockchain have to store the entire blockchain state. This is both a
scalability problem and makes it impractical and time-consuming to add new full nodes to the blockchain.
The work by Bernardini et al.~\cite{bpp-bmdhtdvfss-19} proposes to keep the
blockchain state in an Authenticated Data Structure
(ADS,~\cite{tamassia2003authenticated}) stored in a Distributed Hash Table
inspired to Kademlia~\cite{maymounkov2002kademlia}. The chosen ADS is a
variation of Merkle Hash Tree (MHT,~\cite{merkle2001certified,
merkle2019protocols}) and in particular it is a binary prefix tree equipped with
the same hash linkage of a MHT. In the model adopted
by~\cite{bpp-bmdhtdvfss-19}, the blockchain state is made of the value of the
accounts (one value for each address) and is conceptually stored at the leaves
of this ADS (indexed by address).

Regarding the relation with the DHT, in~\cite{bpp-bmdhtdvfss-19} each blockchain node is
also a Kademlia~\cite{maymounkov2002kademlia} node that stores a pruned version of the ADS, called
\emph{pADS}. As in the regular Kademlia, nodes have a (random) identifier
extracted from the same space of the keys of the data. Here the keys are the
addresses of the accounts. Nodes receive new blocks in broadcast. From each
block, they obtain updated information for all addresses updated by that block.
A node \emph{retains} only those data that are \emph{close} to its node identifier. Here 
closeness is intended in Kademlia sense, that is, according to the well known Kademlia
xor metric~\cite{maymounkov2002kademlia}. Each node stores in its pADS the paths of
the leaves related to addresses retained by the node, up to the root. The pruned parts are those not retained or 
not used by any address. For those pruned parts, a node just keeps the \emph{root hash} of the subtree (see~\cite{bpp-bmdhtdvfss-19}).
The result is a network that keeps the blockchain state as Kademlia does (with replication) but
whose response can be checked against a trusted root hash as for regular MHTs.
All nodes get from the ``last block'' the trusted root hash to use for checking 
that the replies of the Kademlia network are genuine. This check involves
obtaining a \emph{Merkle proof} (i.e., the siblings of the nodes form the leaf to the root in the ADS) for the required value.

We call \emph{block producers} the nodes that create a new block. The considered
model does not constrain the kind of consensus.
In this model, a block producer can perform its task without keeping any state.
This is made possible by the way in which transactions are created. A node that creates
a transaction $t$ asks the Kademlia network for the value and the Merkle proofs
for all the addresses that $t$ is going to change. These Merkle proofs are
attached to the $t$. No signature is required for them so they do not need to be
stored in the blockchain. Merkle proofs are taken by each block producer instead
of accessing a local copy of the state or instead of asking it to the Kademlia
network. With that Merkle proofs, block producers can also reconstruct the part
of the ADS that is affected by each block and compute the next root hash of the
ADS and include it in the produced block.

This machinery is further complicated by the fact that, due to network delays,
not all Kademlia nodes store the blockchain state updated to the same block. To
overcome this problem all nodes that need to handle and check Merkle proofs
stores a queue of the last $l$ blocks from which to take the right root hash
against which to check the replies of the Kademlia network.

In general, transactions arrive to block producers with Merkle proofs that are
late by a certain number of blocks. Block producers have to update the Merkle
proofs to match the root hash declared in the last block before they can be
merged in a pADS. It can be proven that this can always be done securely, even
if the actual algorithm may be tricky.

The primary objectives of our experimentation were to develop a realistic
methodology to estimate the value $l$ and to check in practice that all the
tricky details of the model actually worked, comprising the Merkle proof
updating procedure.

Provided the complexity of the methodology, realizing it in a simulator seemed not
viable to us. Further, realistic delays were needed for the estimation of $l$ and for the 
test of the machinery needed to update the Merkle proofs.
This forced us to head toward emulation instead of simulation. 

We discuss specific problems and our experimental setting in
Section~\ref{sec:experiments-scientific}.

\subsection{Emulation of a Realistic Permissionless Blockchain Network}\label{sec:context-real}

In our second context, we intend to emulate a realistic permissionless
block\-chain. In this experiment, the primary objective of our experimentation is
to create a realistic environment with as many nodes as possible. This kind of
experiment can be useful to software developers to understand how changes or
tuning of the underlying protocols may impact on the large scale behavior of the
whole network. Clearly, the complexity of real software cannot be reproduced in
a simulation and hence emulation is the only viable approach, in this case.
However, dealing with real software has some tricky aspects that may make this
kind of emulation hard.

We can make the following general considerations. For production-ready blockchain
software, it is common to have the following characteristics.

\begin{enumerate}

\item The software is tailored for the typical use case, which, for permissionless blockchains, is ``attaching
one node to an already-existing network''. This might have an impact on how the software
behaves at startup (e.g., regarding neighbor discovery) and on the available configuration
options. 

\item Normally this kind of software is supposed to run on well-equipped hardware.

\item It is designed to run continuously for a long time and possibly, for full
nodes, collecting a huge amount of data.

\end{enumerate}

On the contrary, for our experiment, the following statements hold that
contradicts the above listed typical assumptions.

\begin{enumerate}

\item The network is created all-at-once, since, in the typical experiment run, there
is no preexisting network.

\item Even if the hardware we use to run the emulation can be fairly powerful,
we aim at emulating a large number of nodes, hence, each one is supposed to get
only a small share of the available resources.

\item Each node is supposed to run only for the amount of time strictly needed
to achieve a certain experimental objective. Hence, the amount of blocks and
transactions that need to be processed and stored are quite limited.

\end{enumerate}

The possibility to run an experiment effectively, largely depends on the chosen
technology and to which extent its behavior can be tweaked to match the needs of
our experimental environment. Permissionless blockchains are always based on
open-source software, hence, in principle, any tweak can be implemented by
touching the code. However, it is desirable to avoid this approach: it is
error prone, it may depend on a specific version of the software, it is time
consuming, it might involve non-documented features of the software, etc.

The following are some of the problems that one can expect to encounter.

\begin{itemize}

\item Supposing the experiment does not need any special topology, we would like
to be able to exploit the neighbor discovery of the blockchain. However, this
must work within the closed environment of the emulation, hence, a special
configuration might be needed to achieve this. For example, each node should
be started so that, it points to a specific node (a \emph{boot node}) whose
address is fixed and known to all other blockchain nodes in our
experimentation.

\item If the experiment requires a controlled topology, we should look for
configuration options that allow us to fix the neighbors. The availability of these
options clearly depends on the software implementation.

\item Each node should have enough CPU power to keep the pace of arriving
transactions and blocks. This problem is addressed in
Section~\ref{sec:inflation} with a general approach.

\item Since full blockchain nodes usually are very hungry for persistent storage,
the default configuration of our software usually assumes plenty of it. However,
this is not the case for our emulation. Hence, we expect to have to look for
options that reduce the need for persistent storage at startup.

\end{itemize}

For our test, we choose to emulate a PoS-based Ethereum network
exploiting the very same software that would run on real Ethereum nodes and, in
particular, Go-Ethereum~\cite{geth} plus Prysm~\cite{Prysmati39:online}.

In Section~\ref{sec:experiments-ethereum}, we discuss the problems we encountered and the
configurations that we had to apply in this specific case.

\section{Launching a Large Number of Nodes}\label{sec:launching}

To emulate a large blockchain network, for example, having one Docker container for each node,
we need to be able to launch a very large number of processes. Linux kernel adopts a
defensive approach regarding resource usage. It has a system of limits that
protects the whole system from predatory behavior of certain users or processes
and that can make the whole system unusable. While this is clearly a must-have
feature for multiuser systems to protect against denial of service attacks, in our
case there is no reason to adopt this kind of limits. We are supposing the
machine that we use to run our experimentation is solely dedicated to this purpose.

The first two parameters that limit how many Docker containers we can launch are:
\begin{enumerate*}[(1)]
	\item the number of open files and
	\item the maximum number of processes.
\end{enumerate*} 
In Unix, we have these two limits for each user.
Note that, the Docker hypervisor runs as root, as well as all the processes that are launched within the container.
So we need to adjust these limits for the root user. These limits are controlled by the \emph{ulimit} (\emph{user limit})  settings. There exist two types of ulimit, namely \emph{hard} and \emph{soft} limits. This distinction makes sense in an environment in which a user may need to raise its limit temporarily and autonomously. In our case, we just raise both kinds of limits.
We can permanently change both limits by editing the file \path{/etc/security/limits.conf}. This file is read by the PAM module pam\_limits and applied upon login.
Note that, in general, each blockchain node has multiple open files and user processes. The actual amount depends on the specific 
blockchain software that we intend to run. On the other hand, if we work on a dedicated machine there is no reason to keep these limits low.
In our case, we add the lines \inlinecode{root hard nofile 1574415} and \inlinecode{root soft nofile 1574415} to increase the hard and soft limits concerning the number of open files allowed for the root user. We add the lines \inlinecode{root hard nproc 1574415} and \inlinecode{root soft nproc 1574415} to increase the hard and soft limits concerning the maximum number of processes available for the root user. 

A number of kernel parameters may also affect the scalability of our experiments. These are kernel level 
parameters and affect the system as a whole. Again the main reason for them is to protect the whole system from undesirable resource depletion, and many limits can be arbitrarily raised to ease our experimentation.

While the relevant parameters may depend on the specific needs of each experiment, in Table~\ref{table:param}, we show a list of parameters whose tuning turned out to be useful for our experimentation contexts.
Parameter values are changed by editing the file \path{/etc/sysctl.conf} (applied at boot time), or 
changed and inspected at run time using the \inlinebash{sysctl} command. 
The table shows the name of the parameter, a small description, a default
value (based on the Debian 11 Linux distribution) and an example line to add to the 
\texttt{/etc/sysctl.conf} configuration file.

\begin{table*}[h!]
	\centering
	\begin{adjustbox}{scale=0.7}
		\begin{tabular}{|c|M{0.3\linewidth}|c|I|}
			\hline
			\rowcolor{gray!15}
			\textbf{Parameter}   & \textbf{Description} & \textbf{Default} & \textbf{Example line} \\ \hline
			pty	&  Maximum number of pseduo-terminals. Each Docker container normally uses one of them. & 4096   &          kernel.pty.max = 11000             \\ \hline
			rmem\_max	&         The maximum receive socket buffer size in bytes.  & 212992         &          net.core.rmem\_max=2147483647             \\ \hline
			rmem\_default	&       The default setting of the socket receive buffer in bytes.     & 212992       &         net.core.rmem\_default=2147483647              \\ \hline
			wmem\_max	&        The maximum send socket buffer size in bytes. &      212992        &           net.core.wmem\_max=2147483647             \\ \hline
			wmem\_default	&       The default setting  of the socket send buffer in bytes.        &   212992    &           net.core.wmem\_default=2147483647            \\ \hline
			tcp\_rmem	&        Contains three values that represent the minimum, default and maximum size of the TCP socket receive buffer.        & 4096 131072 6291456
			&     net.ipv4.tcp\_rmem="10240 87380 16777216"                   \\ \hline
			tcp\_wmem	&           Contains three values that represent the minimum, default and maximum size of the TCP socket send buffer.           & 4096 16384 4194304
			&             net.ipv4.tcp\_wmem="10240 87380 16777216"           \\ \hline
			gc\_thresh1	& The minimum number of entries to keep in the ARP cache. The garbage collector will not run if there are fewer than this number of entries in the cache.  & 128 &     net.ipv4.neigh.default.gc\_thresh1=200000                  \\ \hline
			gc\_thresh2	&The soft maximum number of entries to keep in the ARP cache. The garbage collector will allow the number of entries to exceed this for 5 seconds before collection will be performed. & 512 &  net.ipv4.neigh.default.gc\_thresh2=200000               \\ \hline
			gc\_thresh3	&The hard maximum number of entries to keep in the ARP cache. The garbage collector will always run if there are more than this number of entries in the cache. & 1024 &         net.ipv4.neigh.default.gc\_thresh3=200000              \\ \hline
		\end{tabular}
	\end{adjustbox}
        \caption{Table showing some relevant kernel parameters.}
 	\label{table:param}
\end{table*}

\section{Connecting a Large Number of Nodes}\label{sec:connecting}
To connect many blockchain nodes among them, the natural approach is to attach each container to a virtual Linux bridge.
One of the first problems encountered toward scaling to a large number of blockchain nodes is the fact that Linux bridges support a limited number of \emph{ports}, i.e. the virtual interfaces (called \emph{VETH}) among which packets are switched.

In the following, we refer to versions 4.19.208 and 6.1.70 of the Linux kernel.
The maximum number of ports allowed on a Linux bridge is 1024. This value is controlled by a hard coded parameter in the kernel named BR\_MAX\_PORTS, which is defined in the file \path{net/bridge/br_private.h}, as \inlinecode{\#define BR\_MAX\_PORTS  (1 <{}< BR\_PORT\_BITS)}, with BR\_PORT\_BITS immediately defined one line above as \inlinecode{\#define BR\_PORT\_BITS	10}. This means that a default Linux bridge is limited to $ 2^{10} =1024$ ports. If we need to connect $N>1024$ containers, one possibility is to increase BR\_PORT\_BITS to obtain at least $ N \leq  2^{\textrm{BR\_PORT\_BITS}}$ (i.e.,  $N \leq \textrm{BR\_MAX\_PORTS})$.
After this,  we need to recompile the kernel to make the change effective. 
In our case, we set BR\_PORT\_BITS=17.

A second possibility, to get more ports without recompiling the kernel, maybe to use multiple bridges connected to each other. We do not recommend this approach for two reasons: it introduces an additional emulation-level network topology which might be unhandy to manage, and it requires the kernel to perform switching more than once for each packet sent among nodes.

A Linux bridge (like real bridges) is equipped with a 
forwarding database. It stores which level-two MAC addresses 
are present on which virtual interface 
(usually called \emph{VETH}) connected to the bridge.
The forwarding database is populated by a learning process: when an incoming level-two packet is received from 
VETH $v$ with source MAC addresses $a$ a corresponding entry $\langle a, v\rangle$ is recorded in the 
forwarding database. If a packet with destination $d$ has to be forwarded, $a$ is first searched in 
the forwarding database. If an entry $\langle d, w\rangle$ is found, the packet is forwarded on VETH $w$
otherwise, the packet is broadcasted on all VETHs.

At the beginning of our emulation experiment, when nodes start to contact other nodes, the forwarding
database is still empty, hence packets destined to nodes that have not spoken yet are forwarded to all other nodes.
Each of these packets has to be processed by all other nodes (mostly just to be discarded). Supposing to have orders of thousands of 
nodes, this process can easily generate orders of millions of useless packets, making the startup of the 
experiment slow and unreliable (since timeouts are involved in communications among nodes).
To solve this problem, we programmatically initialize the forwarding database with a \emph{static entry} for each
container. We can do that by running,
on the host, the following command for each container of our experiment,
\inlinebash{bridge fdb add <MAC> dev <VETH> master static}. After this
configuration, each packet is regularly forwarded only to the correct node since the beginning.

\section{Addressing the ARP Broadcast Problem} \label{sec:autoarpd}

When a blockchain experimentation involving a large number of nodes is started,
each node starts to talk to other nodes through the bridge that connects all of them. 
Since we aim at realistically emulating networking, 
we intend to support IP and all the above network
stack as occurs in the real internet.
To send IP packets to the right destination (which is known by its IP address), they
have to be encapsulated into level-two frames whose destination addresses are so
called \emph{MAC addresses}. 
All regular IP networks adopt the \emph{Address Resolution Protocol} (ARP) to
obtain a corresponding MAC address when they have to send a packet to a certain IP
address. ARP performs a level-two broadcast to ask who owns a certain IP
address (this is called \emph{ARP request}). The owner of the IP address sends an unicast
\emph{ARP reply}. This occurs the first time a node has to send an IP packet
to a certain destination on the bridged network. The obtained association
called \emph{ARP entry}, is stored in an \emph{ARP cache} and it is used for the
subsequent packets to be sent to the same destination. Actually, each ARP entry
has a further field called \emph{NUD} which will be explained later.

Observe that, when the experimentation starts, each node starts talking with a
number of other nodes. Each of these communications gives rise to a broadcast ARP
request. The kernel duplicates all these requests for all nodes connected to the
bridge (i.e., a quadratic number of communications) and processes each of them when they reach
the destination. This easily increases CPU load to the point that the whole
emulation is greatly delayed. If some sort of timeout is in place (e.g.,
TCP connection timeouts or application specific timeouts), nodes can easily fail
to contact each other.

In real networks, ARP is fundamental to  make management of these
network easily, by dynamically finding MAC addresses corresponding to IP addresses when
needed. In principle, the same result could be obtained by configuring static
ARP entries (which could be configured by using the \inlinecode{arp} command).
A static configuration would completely solve the broadcast ARP problem. However, 
we are now faced with the following two additional problems: \begin{enumerate*}[(1)] \item it is not obvious which node will contact which other since this might be
decided at running time and possibly on a random basis, \item inserting all
possible static ARP entries replicated for each node means to force the kernel to store a quadratic number of ARP entries.
\end{enumerate*}

Regarding the second aspect, the kernel is tuned for small ARP caches and hence
their size is limited. For this reason, the standard tools
that ship with a Linux system provide a well-known user space solution named
\emph{arpd} (whose source code can be found in~\cite{iproute2}).
It is a user-space daemon that helps the kernel in keeping a large ARP cache on
the local disk. It is clearly less efficient than a kernel-only solution but it
scales well when a very large number of ARP entries is needed. We will show
below that arpd turns out not to be a good solution to our problem. However,
since our proposal is derived from arpd, it is useful to understand its design.

Arpd interacts with the kernel using the \emph{NETLINK} protocol~\cite{netlink}
as follows. First, the kernel is instructed not to perform an ARP request\footnote{This is obtained by configuring the following parameter for a specific network interface: mcast\_solicit = 0} but to
ask to the arpd process\footnote{This is obtained by configuring the following parameter for a specific network interface: app\_solicit = 1}.
When a process (i.e., in our case the process implementing a
blockchain node running in a Docker container) intends to send a message to a
certain IP address for which no entry is present in the ARP cache, the kernel
sends a request for that IP address to the arpd daemon (which should be running
in the container). It searches that address in its database and, if it is present, sends
back a response to the kernel, which can be modeled as a triple $\left\langle \textrm{IP addr},
\textrm{MAC addr}, \textrm{NUD} \right\rangle $ and can be put in the ARP cache by the kernel.  The term NUD stands for
\emph{Neighbour Unreachability Detection} and identifies the state of a neighbor
entry in the ARP cache. It can assume many values, comprising the following two
(we omit details that are irrelevant for our context\footnote{A complete
description of the ``neighbouring subsystem'' in Linux can be found
at~\cite{benvenuti2006understanding}}).

\begin{description}
\item[$\textrm{NUD\_REACHABLE}$] It means that the entry was recently used to send an IP packet.
\item[$\textrm{NUD\_STALE}$] It means that the entry was not used recently, hence before being used again a \emph{reachability confirmation} procedure should be performed.
\end{description}

A first solution to our ARP broadcast problem is to initialize the arpd database
with all the entries for our network. However, due to the peculiar way arpd
works, this still leaves in the network a large number of unicast ARP requests.
In fact, unfortunately, the arpd daemon instructs the kernel to insert an ARP
entry labeled $\textrm{NUD\_STALE}$. In this case, the kernel has to perform a
reachability confirmation. This means an unicast ARP request has to be
performed before using that entry. If successful, the entry is labeled as
$\textrm{NUD\_REACHABLE}$ and can be actually used. For this reason, we gave 
up on this approach.

Note that, for the vast majority of blockchain experiments the actual MAC
addresses are not relevant. This enables the possibility to coordinate MAC addresses with IP
addresses to perform this translation without resorting to any protocol.

First note that Docker allows the user to customize the MAC address of virtual
network interfaces. Further, it also has a peculiar approach to set default MAC
addresses when the IP address of the interface is known at the container start up.
For a virtual interface whose IP address is $b_1.b_2.b_3.b_4$, with $b_i$ in $0\dots 255$ ($0\dots \mbox{FF}$ hexadecimal),
the corresponding default mac address for the interface is $02\mbox{:}42\mbox{:}b_1\mbox{:}b_2\mbox{:}b_3\mbox{:}b_4$\footnote{Contrary from regular MAC addresses this MAC address does not have a prefix linked with a vendor. Regarding the $02$ (hexadecimal) value of the first address, the second bit set to 1 means
a \emph{locally administered address}, i.e., no ethernet card has by default a MAC address with this bit set to   1. }. In this context, it is possible, in principle, to provide a MAC to IP translation without the need to rely on 
the ARP protocol. 

We developed a daemon named AutoARPD, whose source code is freely available
on-line~\cite{RomaTreU51:online} that interacts with the Linux kernel as arpd
does, but instead of querying its database (or performing ARP requests on its own), it locally
computes the corresponding MAC address from the IP address according to a configurable pattern. In
this way, we get rid of all the ARP traffic and at the same time we do not need
to have any special support hardcoded in the blockchain node software.

More in detail, when a node needs to contact a certain IP $b_1.b_2.b_3.b_4$ address that has no ARP
entry in the ARP cache,
the kernel sends a NETLINK request for that IP address to the daemon AutoARPD daemon, which 
instructs the kernel to configure the following ARP cache entry 
$$ \left\langle
b_1.b_2.b_3.b_4, 02\mbox{:}42\mbox{:}b_1\mbox{:}b_2\mbox{:}b_3\mbox{:}b_4, \textrm{NUD\_REACHABLE} \right\rangle  .$$

Note that AutoARPD labels the entry as $\textrm{NUD\_REACHABLE}$ so that the
kernel will trust and use that entry without performing any reachability confirmation, which is not needed in 
our controlled environment.

The kernel may switch that entry to $\textrm{NUD\_STALE}$ under low usage conditions, which might lead again to a useless reachability confirmation procedure. To avoid this case, 
we can increase a related threshold to exceed the
duration of our experiment \\(e.g., \inlinecode{base\_reachable\_time\_ms =
72000000}, an interface specific parameter).

To enable AutoARPD,  the procedure is the same as enabling arpd. Clearly
AutoARPD should run in each container alongside the blockchain node software.
Further, the following two kernel parameters should be set for each container:
\\ \inlinecode{net.ipv4.neigh.\emph{interface\_name}.mcast\_solicit = 0} and\\
\inlinecode{net.ipv4.neigh.\emph{interface\_name}.app\_solicit = 1}. \\
This can be done either
during creation, by using the proper Docker option (\inlinecode{$--$sysctl}), or
after the start up of the container (either by directly issuing sysctl commands
or by asking AutoARPD to do that).
To perform this change after startup, the container should run in privileged mode.

\section{Emulating Realistic Internet Delays}\label{sec:latency}

To set up a realistic experiment, it is paramount to also emulate non-ideal
aspects of the internet. In particular, for one of our experimentation contexts,
described in Section~\ref{sec:context-scientific}, it is important to reproduce realistic
network delays. In this section, we describe our approach to emulating realistic
delays for large blockchain experiments.

Zhang et al.~\cite{5280241}, recognized the ``internet delay space'' as an important aspect in the design of global-scale distributed systems.
In their work, they analyze delay measured among thousands of Internet edge networks. 
From these observations, they designed \emph{Delay Space Synthesizer} ($DS^2$).
In the $DS^2$ project page~\cite{Internet68:online}, along with the software, they also published two matrices that represent realistic end-to-end internet delays.
In our experiment, we used \emph{Matrix1} (size: 3997x3997, unit: ms) as input to our system that emulates internet delays. Entries of that matrix represent one-way delays.

First, we remark that the network connecting the nodes of our blockchain is a
virtual and very simple one. Packets are switched among nodes by the kernel. For
this reason, we can take advantage of well known tools provided by the Linux
ecosystem for modifying the network behavior. We used the \emph{Traffic Control}
(\emph{TC}) subsystem~\cite{tldp-tc} that helps in policing, classifying, shaping, and
scheduling network traffic, and the \emph{NFTables} (\emph{NFT}) subsystem~\cite{nftables} that
provides filtering and classification of network packets.

Essentially, NFT is used to create a configuration that marks packets with an
integer that identifies a \emph{class}. This is a \emph{mark} for kernel use only, it
does not affect what is received by the destination. Then TC is used to create a
configuration to queue each packet to a distinct queue (one for each possible
value of the packet mark) corresponding to the class of the packet. These
queues are configured to apply the delay corresponding to that class.

Suppose to have 3000 nodes, the delay matrix has 9M entries. Even supposing to
have symmetric delays (as the $DS^2$ matrices are) handling millions of
classification rules in the kernel is clearly not feasible. We proceed as
follows. First, we approximated delays quantizing them at 10ms steps obtaining
184 different delay values (delays range from zero to about 2 seconds, but
extreme cases are sparse) and hence 184 distinct classes. For each class, we
configure an NFT classification rule as follows. We select all pairs of nodes that
are associated with that class and create a \emph{set} of \emph{unordered pairs} of
IP addresses. NFT sets are a kind of data structure supported by the Linux netfilter
module that allows us to efficiently match a packet against a large set of
addresses, or pair of addresses as in our case.

An integer mark is assigned to each class as follows.
Classes are sorted according to their associated delay, from small to large delays. 
The mark of a class is the position of that class in this order, starting from 1. 

The creation of the NFT configuration is detailed in
Algorithm~\ref{algo:delayClassification}. This algorithm takes as input a map
that associates for each mark, which identifies a class, its set of IP address
pairs and produces as side-effect the NFT configuration.

\begin{algorithm}[t]
	\caption{Procedure CreateNFTConfiguration creates netfilter rules to mark each packet with its delay class.}
	\label{algo:delayClassification}
	\begin{algorithmic}[1]

		\Procedure{CreateNFTConfiguration}{classes} 
		
		\State \textbf{Input:} classes: a map from a numeric mark that identifies a class to a set of pairs of IP addresses.
		
		\State \textbf{Side Effect:} NFT is configured with a new table that marks each packet with the mark associated with the correct delay class on the basis of the source and destination addresses.
		\LineComment{Create a table with name \emph{latem} with address family \emph{ip}}
		\StateBash{nft add table ip latem}
		\LineComment{Add a chain with name \emph{latem\_chain} in \emph{latem} }
		\StateBash{nft add chain latem latem\_chain \{ type filter hook forward priority 0 $\backslash$; \}}
		\ForAll { pairs $(m,  \mathrm{pairSet} )$ in classes}
		\LineComment{Declare in NFT a set of type
		$ \{\textrm{ipv4\_addr}\ . \ \textrm{ipv4\_addr} \} $ 
		 with name \emph{nodes\_}$\langle m\rangle$.  
		The dot ``.'' is used by NFT to separate source and destination in a pair.}
		\StateBash{nft add set latem nodes\_$\langle m\rangle$ \{ type ipv4\_addr . ipv4\_addr $\backslash$; \}}
		\State $N \gets \emptyset$
		\ForAll{ $(s, d)$ in pairSet }
		\State Add \textquotesingle\textquotesingle$ s $ . $ d $\textquotesingle\textquotesingle  ~in $ N $
		\State Add \textquotesingle\textquotesingle$ d $ . $ s $\textquotesingle\textquotesingle  ~in $ N $
		\EndFor
		\LineComment{Add all element to the set}
		\StateBash{nft add element latem nodes\_$\langle m\rangle$ \{ $\langle \mathrm{N}\rangle$\ \}}
		\LineComment{add a rule to mark a packet matching $N$ }
		\StateBash{nft add rule latem latem\_chain ip saddr . ip daddr @nodes\_$\langle m\rangle$ meta mark set $\langle m\rangle$}
		\EndFor
		\EndProcedure

			\end{algorithmic}
\end{algorithm}

\begin{algorithm}
	\caption{Procedure CreateTCConfiguration creates the TC configuration to reproduce the tree shown in Figure~\ref{fig:tcTree} to apply the delay at each packet according to the class identified by its mark.}

	\label{algo:delayApplication}

	\begin{algorithmic}[1]
				\Procedure{CreateTCConfiguration}{classDelays, $ v $, $ b $}
		\State \textbf{Input:} classDelays: map from mark values to the delay associated with the class identified by that mark.
		\State \phantom{\textbf{Input:}} $ v $: name of a virtual interface to apply TC configuration to
		\State \phantom{\textbf{Input:}} $ b $: maximum number of bands in each prio qdisc
		\LineComment{Create the root prio qdisc for $v$ with $b$ bands.}
		\StateBash{tc qdisc add dev $\langle v\rangle$ root handle 1: prio bands $\langle b\rangle$}
		\LineComment{Create the second level of prio qdiscs, each attached to a different band of the root and each having $b$ bands.}
		\ForAll{$ i $ form $ 1 $ to $ b $}
				\StateBash{tc qdisc add dev $\langle v\rangle$  parent 1:$\langle \textrm{hex}(i)\rangle$ handle 1$\langle \textrm{hex}(i)\rangle$: prio bands $\langle b\rangle$}
		\EndFor
		
		\LineComment{create all netem qdiscs with proper delay and their corresponding filters}
		\ForAll{$( m, d )$ in  classDelays}
		\State $ f \gets  \floor*{\frac{m-1}{b}}+1$ \Comment First level band. 
		\State $ s \gets ((m-1)\  \textrm{mod}\  b ) +1$  \Comment Second level band.
		\LineComment{create the netem qdisc with the correct delay}
		\StateBash{tc qdisc add dev $\langle v\rangle$ parent 1$\langle\textrm{hex}(f)\rangle$:$\langle\textrm{hex}(s)\rangle$ netem delay $\langle d \rangle$ms}
		\LineComment{create a path of filters from root to the just created netem qdisc}
		\StateBash{tc filter add dev $\langle v\rangle$ protocol ip parent 1: } 
		\StateBash{ \hskip 2cmprio 10 handle $\langle m\rangle$ fw classid 1:$\langle\textrm{hex}(f)\rangle$}
		\StateBash{tc filter add dev $\langle v\rangle$ protocol ip parent 1$\langle\textrm{hex}(f)\rangle$:}
		\StateBash{ \hskip 2cm prio 10 handle $\langle m\rangle$ fw classid 1$\langle\textrm{hex}(f)\rangle$:$\langle\textrm{hex}(s)\rangle$}
		\EndFor

		\LineComment{Create low priority filters to match all unmatched packets. They correspond on the rightmost path in Figure~\ref{fig:tcTree}. No netem qdisc is needed here since no we do not apply any delay.}
		\StateBash{tc filter add dev $\langle v\rangle$  protocol all parent 1:} 
		\StateBash{ \hskip 2cm prio 20 matchall classid 1:$\langle\textrm{hex}(b)\rangle$}
		\StateBash{tc filter add dev $\langle v\rangle$  protocol all parent 1$\langle\textrm{hex}(b)\rangle$:}
		\StateBash{ \hskip 2cm prio 20 matchall classid 1$\langle\textrm{hex}(b)\rangle$:$\langle\textrm{hex}(b)\rangle$}

		\EndProcedure
	\end{algorithmic}

\end{algorithm}

\begin{figure}
	
	\centering
	\includegraphics[width=.9\linewidth]{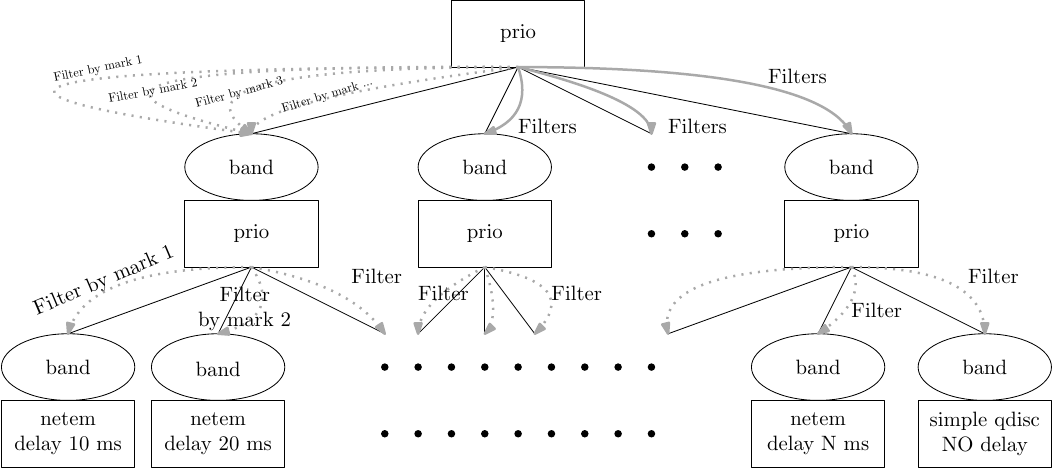}
	\caption{TC tree with two levels of prio qdiscs. The rightmost path is associated with 
	non-marked packets that do not have to be delayed.}
	\label{fig:tcTree}
\end{figure}

Each class of packets has to be treated differently, and in particular, their
packets have to wait in a queue for the delay time associated with their class. We
can do this by using TC. TC involves the use of \emph{qdiscs} (\emph{queuing discipline}) and \emph{filters} to
create a tree-like structure, where the root is the queue where \emph{incoming} or \emph{outgoing} packets are put. Afterward, filters sort the packets into the
various branches until they reach the leaves of the structure, where a final
queue represents the final destination queue of the packet. 
In our experiment, we (arbitrarily) chose  to delay packets outgoing from the bridge,
adopting the same filtering/qdisc structure for all 
network interfaces, which is depicted in Figure~\ref{fig:tcTree}.
The semantics of the elements are as follows.
\begin{description}

	\item[\textbf{prio}] is a \emph{classful} qdisc that is able to dispatch
	packets into an array of \emph{bands} (a sort of ``channels'', at most 16 bands
	are available). Which bands is selected for each packet depends on the
	\emph{filter} attached to each band. When a packet is dispatched to that band it can be processed by another qdisc. The original purpose of the prio qdisc is to prioritize traffic. However, we
	essentially use it to provide a first level of sorting on the basis of the
	mark attached to the packet. We use them as internal nodes of the tree.

	\item[\textbf{filter}] is a rule attached to a band of a classful qdisc to
	determine if a packet has to be dispatched into that band. We use them to
	select the packet to the correct band of a prio qdisc  on the basis
	of the mark of the packet.

	\item[\textbf{netem}] is a classless qdisc that is used to add delay, packet
	loss, duplication and other characteristics to packets. We use them as leaves
	of the tree. In our experiments, we used only the delay feature.
\end{description}

The main idea is to define a tree with a number of leaves equal to the number of
delay classes to be emulated plus one additional leaf. This additional leaf is configured with
no delay and it is used for traffic with no mark attached. In our setting, this traffic is only 
the one generated by communication between each node and the host.

Since we have more than 16 classes, we have to perform two sorting levels (see
Figure~\ref{fig:tcTree}). In the first level, filters match ranges of classes.
In the second level, filters match at a finer granularity only within the range
selected in the first level and select the netem qdisc that will apply the
correct delay. With this scheme, we can support up to 255 delay classes.
Algorithm~\ref{algo:delayApplication} shows the procedure
\emph{CreateTCConfiguration}, which configures TC to create the above described tree structure.
The rightmost path of the tree is associated with the default no-delay leaf of the tree.

Algorithm~\ref{algo:delayApplication} takes as input the delay to be applied to
each class. This is represented as a map from a mark value to the delay value
for the class identified by that mark. It also takes as input the name $v$ of the
VETH to which the configuration has to be applied. We intend that
Algorithm~\ref{algo:delayApplication} is run for all the VETHs of the
bridge (by default the configuration is applied to the traffic outgoing from the bridge). It also
takes a parameter $b$, which is the maximum number of children of the node of
the tree (i.e., the number of bands for all prio qdiscs in the tree). In our case,  it should satisfy the following inequality $b^2 \geq
184+1$ (i.e., all classes plus the default one for non marked packets), which
gives $b=14$. In the algorithm, the following notations imposed by the
\texttt{tc} command syntax~\cite{man-tc} are used to refer to elements of the tree. Each qdisc
(prio or netem, see Figure~\ref{fig:tcTree}) is identified by a \emph{handle},
which is an arbitrary hex number followed by ``:'' (this is the notation used in
the \texttt{tc} commands). We denote by $\mathrm{hex}(y)$ the hex representation
of number $y$. By convention ``\texttt{1:}'' is the handle of the root. Bands
are identified by $h$\texttt{:}$c$, where $h$ is the handle of the qdisc this
band is related (i.e. the parent of the band) while $c$ identifies the band
among those of the parent ($c$ starts from 1). When a qdisc is attached to a
band (called the parent of the new qdisc) the band identifier should be
specified as parent according to the \texttt{tc} syntax.

We note that, by introducing artificial delays we also affect CPU load.
For an experiment with no artificial delays configured, broadcasts are ideally
propagated instantaneously and CPU bursts for processing propagated transactions
or blocks are all close together. In this case, the propagation speed is
dominated by the speed of the CPU, which is the bottleneck of the system in that
specific instant of time. On the contrary, in an experiment with artificial
delays, the CPU load related to a block or a transaction is more evenly
distributed over time and the system is not hindered by a CPU limit, but only by the
configured delays. We artificially exploit and tune this phenomenon in
Section~\ref{sec:inflation}.

\section{Time Inflation}\label{sec:inflation}

Every physical machine has limited resources. Essentially, the bottleneck that
limits the number of nodes of our experimentation can either be the CPU or the amount of
RAM. In principle, it should be possible to run the emulation at the highest possible speed 
so that the CPUs are never idle. However, this approach would make the timings of our emulation unrealistic. Hence, we aim not to have overloaded CPUs so that to obtain realistic timings. 
Up to a certain extent, the approach shown in Section~\ref{sec:latency} allows us to reach that goal. However, increasing the number of nodes, at a certain point, CPUs turn out to be overloaded again.

In this section, we show  a technique, which we call \emph{time inflation}, which is extremely useful for working around CPU limits at the price of 
an increased duration of the experiment.

The main idea is to inflate by a factor $ x $ all the delays involved in our experimentation. 
The net effect is to increase by a factor $x$ also the duration of the experiment. On the other hand, 
this has also the effect of letting the experiment run as if the CPU speed was also increased by a factor $x$. 

To correctly apply this technique, we have to perform the following three kinds of inflation.
\begin{enumerate}
\item Network delays. In our case, this is very simple: we just increase by a factor $ x $ all the delays in \emph{Matrix1} (see Section~\ref{sec:latency}).
\item Timers involved in the execution of the blockchain node software. For a blockchain, the most important of them is likely to be the block time. This was easy in our case since the software of the node was written for the experiment. When using production software, timers are likely to be configurable. 
\item Timers involved in the generation of the load. If the objective of the experiment is to show the possibility of processing a certain transaction load, this should be also spread over time by a factor $x$. 
\item Timers involved in the execution of protocols that are under the control of the kernel. These are essentially timers related to TCP. 
\end{enumerate} 

While for the first three items, there is not very much to discuss, the last
aspect is quite critical. In fact, increasing the delays of the network,
TCP packets can exceed the TCP \emph{retransmission timeout}
(\emph{RTO}). This means that the sender does not receive the ACK regarding the
sent packet before the timeout expires, even if the packet is not lost and it is correctly
delivered. Hence, the sender mistakenly detects a packet loss and retransmits the
packet, causing an artificial increase in the network load (and hence of the CPU load).

The retransmission timeout is dynamically managed\footnote{Exponentially increasing at each retransmission.} by the kernel on a per-connection basis.
Unfortunately, the initial RTO timeout is a value hard coded in the kernel, 1 second in the kernel version we used, which implies that during the TCP three-way handshake, spurious retransmission occurs if one way delays are larger the 0.5 seconds.
The easiest way to change this initial RTO value is to recompile the kernel. However, in an experimental setting, it is quite unhandy to recompile the kernel every time we want to increase or decrease this parameter to match the time inflation. 

We have adopted a more comfortable solution to modify the initial RTO in a
flexible way based on the \emph{Berkeley Packet Filters}  (\emph{BPF}) facility
of the Linux kernel. This tool allows us to write code that is ``attached'' to a
designated code path in the kernel. When the code path is traversed in the processing of the packet, the
attached BPF program is executed. The BPF is quite efficient since code is
compiled. Further, in our case, the BPF code is executed only during connection
establishment to correctly set the initial value of the RTO of the new
connection. In the following, we provide some details on how to realize this
solution.

The initial RTO value is defined in the file \path{include/net/tcp.h} of the Linux
kernel source code. It is defined as \inlinecode{\#define TCP\_TIMEOUT\_INIT
((unsigned)(1*HZ))}, where \inlinecode{HZ} is a constant stating the quantity of
certain interrupts that the kernel performs per second\footnote{For our system, HZ=250. It can be found with the following command: \inlinebash{grep \textquotesingle CONFIG\_HZ=\textquotesingle 
 //boot/config-\$(uname -r)}}~\cite{time7Linux, Linuxman}. Hence, \inlinecode{1*HZ} means 1 second.
The default RTO value is computed by a specific Linux kernel function~\cite{tcphincl}, which provides a hook for optional BPF code, and defaults to returning \inlinecode{TCP\_TIMEOUT\_INIT}. The BPF code in Figure~\ref{fig:BPF-RTO}
exploits that hook. This is a C code that should be compiled with specific compilers and options to produce a BPF-compatible bytecode.
Then, the bytecode should be loaded into the kernel. The bytecode will be compiled in native machine language by the kernel itself.  
The following is the detailed procedure to compile and load into the kernel, the code in Figure~\ref{fig:BPF-RTO}. Suppose the code is in the file \inlinecode{tcp-rto.c}.
\begin{enumerate}
	
	\item To compile the code,  use the following command\\
	\inlinebash{clang -O2 -target bpf -c tcp-rto.c -o tcp-rto.o}.
	
	\item To load the code into the kernel, use the following command\\
	\inlinebash{bpftool prog load tcp-rto.o /sys/fs/bpf/tcp-rto}.
	
	\item Find its \emph{program ID} using the command \inlinebash{bpftool prog show}. We obtain an output like the following\\
	\tikz \node[rounded corners=1mm, fill=gray!15,align=left,node font={\ttfamily \scriptsize}]{
		...\\
		169: sock\_ops  name set\_initial\_rto  tag e4384b8da577553a  gpl\\
		\hspace{10mm}loaded\_at 2021-04-29T15:49:03+0800  uid 0\\
		\hspace{10mm}xlated 296B  jited 186B  memlock 4096B
	};\\
	where the program ID is the number on the right (169 in this example).
	
	\item The BPF program should be also attached to a cgroup using the following command\\
	\inlinebash{\scriptsize{bpftool cgroup attach /sys/fs/cgroup sock\_ops id 169}}.
	
\end{enumerate}

The BPF program can be unloaded by calling the following commands\\
\inlinebash{rm /sys/fs/bpf/tcp-rto}\\ \inlinebash{\scriptsize{bpftool cgroup detach /sys/fs/cgroup sock\_ops id 169}}.

\begin{figure}
\begin{lstlisting}[		
	frame=tb,
	language=c,
	aboveskip=5mm,
	belowskip=5mm,
	showstringspaces=false,
	basicstyle={\scriptsize\ttfamily\bfseries},
	numbers=left,
	numbersep=1mm,
	numberstyle=\tiny,
	keywordstyle={\color{OliveGreen}\normalfont\ttfamily},
	stringstyle=\color{GreenYellow},
	escapechar=|
	]
	#include <linux/bpf.h>
	
	#ifndef __section
	# define __section(NAME)     \
	__attribute__((section(NAME), used))
	#endif
	
	
	__section("sockops")
	int set_initial_rto(struct bpf_sock_ops *skops)
	{ |\label{line:begin}|
		const int timeout = 3; // initial RTO timout in seconds |\label{line:inf}|
		const int hz = 250;  // this value has to match the HZ value of the system |\label{line:hz}|
		
		int op = (int) skops->op;
		if (op == BPF_SOCK_OPS_TIMEOUT_INIT) {
			skops->reply =  timeout * hz; 
			return 1;
		}
		
		return 1; 
	} |\label{line:end}|
	
	char _license[] __section("license") = "GPL";
\end{lstlisting}
\caption{BPF code to provide a custom initialization for TCP retransmission timeout. This allows us to correctly handle TCP connection setup in large scale blockchain experimentation with inflated time without recompiling the kernel.}
\label{fig:BPF-RTO}
\end{figure}

While the BPF technology imposes quite a lot of boilerplate code, the important part of the code shown in Figure~\ref{fig:BPF-RTO} is quite simple (Lines~\ref{line:begin}-\ref{line:end}).
Line~\ref{line:hz} should be changed to match the HZ of the system and Line~\ref{line:inf} to match our desired timeout in seconds (e.g. 3 seconds in the example).

\section{Experimental Results}\label{sec:experiments}

In this section, we describe the design and execution of the emulation experiments
related to the two contexts described in Section~\ref{sec:context}.

For both experiments, we adopted a single virtual machine configured with 40
cores and 400GB of RAM running Linux as guest operating system. The underlying
physical machine was equipped with two Intel(R) Xeon(R) Gold 6238R CPU @ 2.20GHz
for a total of 112 cores, 1TB of RAM, and two SSDs\footnote{The SSDs are marked
DELL but turn out to be Kioxia KPM6XRUG960G, with SAS interface, 12Gbps bus
speed, 595K IOPS sustained 4KiB random read, 75K IOPS sustained 4KiB random
write.} with RAID~0 configuration. The physical machine runs Windows as
an operating system and HyperV as virtualization software. We were not able to
reserve more than 400GB of RAM for our virtual machine due to the limits of HyperV.
We instructed HyperV to reserve all the RAM of our virtual machine since the
beginning so that no delay is introduced by HyperV when a large amount of RAM is
requested.

As a guest operating system, in the scientific context, we used 
Debian 11 with kernel ver. 4.19.208, while in the 
realistic context, we used Debian 12 with kernel ver. 6.1.70. We used the Docker version
24.0.7.

In both contexts, each of our containers runs a Python script, called
\texttt{node.py}, which helps to start the processes in each container,
incrementally. In both contexts, the first thing that is run is the AutoARPD
software described in Section~\ref{sec:autoarpd}.

The scripts and instructions for reproducing the experimental emulation shown in Section~\ref{sec:experiments-ethereum} are provided in a companion GitHub repository~\cite{companion-repo}.

\subsection{Results for the Scientific Context}\label{sec:experiments-scientific}
By adopting the techniques shown in this paper, we were able to run our scientific blockchain
experiment with 3500 Docker containers, with
realistic network delays. The objectives and scientific context of experimentation were described in
Section~\ref{sec:context-scientific}. 

In our specific experiment, containers are not one-to-one with nodes and nodes may have several different roles. Essentially, each node may take 
from two to four containers. We do not get into further details that are not
very relevant for the purpose of this paper.

The overlay network topology was randomly generated. Our experiment implies the creation
of inter-node connections: the TCP-based are 8000 (for block propagation) and the UDP-based are 64000 (for transaction propagation and the Kademlia protocol). We have a block time of 5 seconds and a load of 20 transactions per block on
average. We set the time inflation factor at $\times 4$ using the approach described in Section~\ref{sec:inflation}.

We run one of our experiments for about 12 minutes of wall-clock time
corresponding to 3 minutes of emulated time. The occupied RAM is about 350GB
which is close to the limits of our virtual machine,  while CPU is below 50\%
on average. Hence, in our setting, the bottleneck is the RAM while without
adopting any time inflation, it would have been the CPU.

Starting up the whole experiment takes quite some time: about 3.5 hours. 
To startup all the needed containers, it takes about 1 hour. Our \texttt{node.py} script runs in each container. It first starts
AutoARPD and waits for a Unix signal before doing anything else. In fact, to
complete the network setup, all the containers have to be up, but it is meaningless to start the experimentation before the network setup is completed. For this reason, 
containers have to be stopped or put in a sleeping state in some way.

To setup configurations for delay emulation and bridge configuration, we need to
gather information about the interfaces of each container. This information
gathering takes about 30 minutes. Setting up static entries for the bridge
forwarding database for all containers takes also about 30 minutes. We think that
these very large times are mostly due to the locking of involved data structures
which essentially impose serialized access. This time is largely reduced with the adoption of a 
newer Linux kernel (see Section~\ref{sec:experiments-ethereum}).

Computing and setting of delay-related configuration, according to what we have described in
Section~\ref{sec:latency}, takes about 40 minutes.

In our experiment, we use two kinds of overlay networks. In both cases, we do
not leave nodes to perform node discovery autonomously, but we configure 
artificially created overlay network routing tables.  
At this point, we trigger the nodes to contact neighbors, which involves setting up about 8000
TCP-based connections and about 64000 UDP-based connections. This takes about 50
minutes. We were surprised to observe that this phase is quite demanding in terms of CPU. We were
able to perform it by sequentially triggering each container to set up these
connections, delaying each trigger by about half a second. 

After this, we trigger the nodes to start working. In particular, certain nodes
are dedicated to creating the transaction load of the network. These nodes are
started after that other nodes are triggered to accept transactions. 
This part takes negligible time.

To orchestrate our experiment, we did not use any special tools. We used a makefile
with \emph{targets} dedicated to startup the containers, to shut them down at
the end of the experiment, to invoke an external Python script for the more
complex settings (see below), to trigger the nodes to start working (using Unix
signals), to activate some logging (for debugging purposes), to stop the
activity of the nodes, and to trigger the nodes to dump specific information
needed for the experimentation. Regarding the external Python script, it takes
care of all the settings concerning the virtual and overlay networks, i.e.,
configuration of static entries on the bridge, creation of the random
topologies for overlay networks, communicating them to the nodes, computation of the delays for the virtual network, and
application of delay-related network configurations.

\subsection{Results for the Realistic Context (PoS-Based Ethereum Network)}\label{sec:experiments-ethereum}

In our realistic context, we set up an \emph{Ethereum 2} network, which is a
PoS-based blockchain. We refer to the old PoW-based Ethereum technology as 
\emph{Ethereum 1}. 
In this section, we \begin{inparaenum}[(1)]\item review the architecture of a node of an Ethereum 2 network (which is important to understand the startup procedure),  \item we provide some details about the specific configuration adopted in our experimentation, 
\item we describe the startup procedure and \item we discuss the RAM occupation, which turns out to be the limiting aspect in our experiment, as desired.
\end{inparaenum}  

Each Ethereum 2 node is realized by running a suitable
compositions of daemons, which are usually grouped into two layers.

\begin{description}

\item[Execution layer.] The main role of this layer is to execute transactions
and run smart contracts in the \emph{Ethereum Virtual Machine}. In our experimentation, this layer
does not execute the consensus protocol, and it is 
realized by running the Go-Ethereum software, which is called
\texttt{geth} (ver. 1.13.8-stable)~\cite{geth}. The running instance of
\texttt{geth},  is also called \emph{execution-node} and usually abbreviated as
\emph{enode}\footnote{ In the old PoW-based
Ethereum 1 technology, a node was made only of this layer, which was in charge of every
activity, including the PoW consensus. With the adoption of the PoS consensus, just the unchanged functionalities are kept
active in this layer and new ones (beacon-chain and PoS consensus) are
implemented in the \emph{consensus layer}~\cite{prysmdocumentation}.}.

\item[Consensus layer] The role of this layer is to participate in a so-called
\emph{beacon-network}, whose nodes keep a blockchain called \emph{beacon-chain}.
The process in charge of realizing this layer is called \emph{beacon-node}. In
our experimentation, to realize this layer, we adopt the \emph{prysm} software
(ver. 4.1.1)~\cite{prysm}, which consists of two daemons:
\texttt{beacon-chain}\footnote{Unfortunately, terminology is quite confusing:
the piece of software that realizes a beacon-node is actually called
\texttt{beacon-chain} in the software package. } and \texttt{validator}. The
first one is the beacon-node, and the second one takes care of staking and block
validation aspects, participating in the PoS consensus process to create new
blocks.
\end{description}

The interaction of these two layers in the Ethereum 2 architecture is very
peculiar. A beacon-node needs to connect to an enode to work, hence, usually an
Ethereum 2 node is formed of one enode and one beacon-node (plus, optionally, a
validator). Enodes and beacon-nodes participate to two distinct and unrelated
peer-to-peer networks. In an Ethereum 2 network, there is a particular instant of
time called \emph{the merge}. Before the merge, enodes realized a traditional
Ethereum 1 PoW-based blockchain network, while beacon-nodes realize a PoS-based
blockchain network whose blocks do not contain any transaction. After the merge,
the Ethereum 1 PoW-based blockchain network stops working and the corresponding
blockchain is now grown by the beacon-network: consensus on new blocks is made by the PoS 
running in the beacon-network and new accepted Ethereum 1 blocks are incorporated as 
part of each block of the beacon-chain (the so-called \emph{execution
payload}\cite{beaconchaincontent} of that blocks). 

We configured the daemons so that the merge occurs immediately. The result is an Ethereum 2 network 
with just a very specific startup process.

After the merge the responsibility of each
part of the Ethereum architecture is the following.

\begin{itemize}
\item Each enode propagates candidate transactions and collects received candidate
transactions in the \emph{mempool}. It serves as the consensus layer to 
validate transactions for received blocks. Additionally, 
in the case of nodes equipped with a validator, it
picks transactions for a new candidate block and validates received candidate blocks for PoS consensus.

\item Each beacon-node receives and propagates new blocks and delegates to its enode 
any validation activity regarding received blocks.

\item Each validator participates in the PoS process. When acting as leader, it
asks to its enode to pick a valid set of transactions from the mempool to
insert in the execution payload of a new candidate block. When voting for a received candidate block, it
asks to the enode to check the validity of the transactions of the execution
payload in the block just received.

\end{itemize}

Our experiment comprises 750 Ethereum 2 nodes. Each node runs \texttt{geth} and
\texttt{beacon-chain} processes, while 20\% of the nodes also run a
\texttt{validator} process. Each node first starts the
\texttt{node.py} python script, which controls the execution of the instances of AutoARPD
(see Section~\ref{sec:autoarpd}), \texttt{geth}, \texttt{beacon-chain}, and
possibly \texttt{validator}.

We intended to experiment using the block time of 12 seconds adopted by the real
production Ethereum 2 network. However, since the CPU load turned out to be too
high, we applied a $\times 2$ time inflation factor to lower it. To do this, we
also double the block time of the beacon-node by proper configuration
(tweaking the \texttt{SECONDS\_PER\_SLOT} parameter in the configuration of the
\texttt{beacon-chain} software).

As already discussed in Section~\ref{sec:context-real}, production-ready
software forces us to cope with certain typical features that may conflict with
our emulation setting. The following are those that we have to face in our case.

\begin{itemize}

\item For each node, we create keys and identifiers in advance (for both
enodes and beacon-nodes processes)  since they are needed in the daemon
configurations to specify neighbors and the set of validators. 

\item We create the genesis block for both the Ethereum 1 chain and the
beacon-chain. Additionally, the genesis block of the beacon-chain contains the
set of validators, which, in our experiment, does not change during the
experiment execution. Here we use the validator identifiers that we compute in
advance.

\item In our experiment, we explicitly set the overlay topologies of the
enodes peer-to-peer network and of the beacon-network. Here we use the
identifiers of the beacon-nodes and of the enodes that we computed in advance.

\item We realize that the default configuration of \texttt{geth} needs
a lot of persistent storage (2GiB multiplied by the number of nodes),
which is not available in our emulation platform. We configured \texttt{geth} to be much
less eager for storage by using the \texttt{--cache} option with a suitable value.
\end{itemize}

In our experiment, the two overlay topologies are generated to be
``small-world''. They are generated using the networkx python library (using the
"newman\_watts\_strogatz\_graph()" function). However, we believe that
topologies have a very limited impact on our experimentation and its feasibility.

The start procedure is quite similar to the one used in the scientific
experimentation described in Section~\ref{sec:experiments-scientific}. It relies
on a bunch of scripts consolidated within a makefile. 

We noted that the adoption of a newer kernel version (ver.~6.1.70 in this case)
obtained us a substantial speedup. This is true especially for the steps where kernel is more
involved, that is, when launching the containers and when configuring the
internet-like delays (see Section~\ref{sec:latency}).

We now detail all the steps of the start up process.

\begin{enumerate}
\item Initially we create all the keys and identifiers for all enodes and beacon-nodes, which 
takes about 1 minute. This also comprises the creation of the Ethereum 1 genesis
block with initialized corresponding account balances. The genesis block is
created to start as PoW chain and arrive immediately to its \emph{terminal total
difficulty} when the merge occurs. This genesis block also includes the
creation of the smart contract used for validator staking.

\item We create directories for each node. This occurs before starting the
nodes themselves, since corresponding directories are mounted into the
containers that realize the nodes. The genesis block for the beacon-chain
is also created in this step. This takes 38 seconds.

\item We start the 750 containers (one for each node) using the -j option of the
make utility without specifying any maximum number of jobs. To start all
containers it takes about 2 minutes (without the -j option it takes 7 minutes).
Each container just starts the \texttt{node.py} script, which in turn
immediately runs AutoARPD and waits for a signal before proceeding. 

\item We complete the setup of the network, in a similar way to what we did in
the scientific experimentation. Additionally, we also create the configuration of
enodes containing network ID, the ``sync mode full'' option, the ``no node
discovery'' option, and, most importantly for us, the specification of static
neighbors. The random topologies for the two peer-to-peer networks are also computed
in this phase. This step takes about 72 seconds, totally.

\item A signal is sent to all nodes to continue the boot. This makes the \texttt{node.py} script of each node to  start \texttt{geth}
(to realize an enode) and then wait for a signal to arrive. Since \texttt{}
immediately transition to Ethereum 2 mode, no PoW is performed, and all \texttt{geth}
instances set up their peer-to-peer network and just wait for a
\texttt{beacon-chain} process to connect.

\item Again a signal is sent to all nodes to run the \texttt{beacon-chain}
process and wait for a further signal. At this point each \texttt{beacon-chain}
process is connected with the corresponding \texttt{geth} process, and the
peer-to-peer network for \texttt{beacon-chain} is established, however, no
blocks are produced since no validator is active at the moment in any node.

\item Again a signal is sent to all nodes, and some of them (20\%)
launch the \texttt{validator} process and wait for a further signal. At this time 
the blockchain starts producing blocks. However, they are still empty since no 
transaction has been produced, yet.

\item When a further signal is sent to all nodes, the production of
transactions is triggered. Transactions production is managed by the \texttt{node.py} script itself. 
Transactions are produced randomly at an expected frequency
that is the same for all nodes. Totally, the frequency of generated transactions
is 4 for each second, resulting in a block of about 50 transactions every 12
seconds. Since in our experimentation, we have configured a $\times 2$ inflation time, we
actually generate 2 transactions for each second and a block every 24 seconds. 

\end{enumerate}

\begin{table}
\begin{tabular}{|l|r|r|r|r|r|}
\hline
\multirow{2}{3.5cm}{} & \multicolumn{3}{c|}{\makecell{Occupation of \\a single node}} & \multicolumn{2}{c|}{Total occupation} \\
\cline{2-6}
 & \makecell{min \\(MiB)} & \makecell{max \\(MiB)} & \makecell{avg \\(MiB)} & \makecell{total \\(MiB)} & \makecell{\% over \\available\\ RAM \\(384GiB)}\\
\hline
\makecell[l]{1. After starting the \\container, only \\ \texttt{node.py}  and \\AutoxARPD running} & 23.06 & 34.09 & 28.9262 & 21694.65 & 5.51\% \\
\hline
\makecell[l]{2. After geth started} & 314.3 & 347.8 & 327.211 & 245408.25 & 62.38\% \\
\hline
\makecell[l]{3. After beacon started} & 396.6 & 435.2 & 412.807 & 309605.25 & 78.70\% \\
\hline
\makecell[l]{4. After validator \\started} & 398.4 & 451.4 & 420.5 & 315375 & 80.16\% \\
\hline
\makecell[l]{5. After processing \\ some transactions} & 255.1 & 342.3 & 286.216 & 214662 & 54.56\% \\
\hline
\end{tabular}
\caption{Memory occupation for the experiment in the realistic context (Ethereum
2 PoS-based blockchain): evolution of the memory occupation through all the
steps of the emulation startup described in the text.}
\label{tbl:memory-realistic-experiment}
\end{table}

As mentioned at the beginning of this section, we aim at running as many nodes
as possible. Since we artificially reduced the CPU consumption by time inflation
and the storage consumption by proper configuration, the bottleneck with respect
to the scalability of the number of nodes is the RAM. In this experimentation
context, the RAM occupation during startup has a peculiar evolution. We
collected the memory occupation of each node by running the \texttt{docker stats}
command on the host at specific points during the startup process.
Table~\ref{tbl:memory-realistic-experiment} shows some summary statistics of our
measurements. The maximum memory occupation occurs when all processes are
started, but no transaction has been produced yet. We presume this behavior is
due to the fact that \texttt{geth} does not free the resources it allocates to
work in ``PoW mode'' until the first validation activity is requested by the
\texttt{beacon-chain} process. In this situation, each node occupies about
420MiB, for a total of about 308GiB of RAM occupied by all 750 nodes. The
\texttt{docker stats} command shows a total of 384GiB available for containers,
hence, at this phase of the startup process, our experiment occupies about 80\%
of the RAM available for containers. We avoided pushing our experiment above
80\% of memory occupation to leave enough room for buffers and caching.

When the first block is accepted, the merge occurs and the behavior of
\texttt{geth} changes considerably. In this new configuration, each node
occupies about 286MiB on average for a total of 210GiB, which accounts for only
54\% of the available RAM. 

Clearly, this RAM occupancy behavior is specific to the tested technology and we
expect other technologies to have a simpler behavior. 

It is worth mentioning that even with this strange behavior, it is still
possible to run additional nodes to exploit more than 54\% of the available
RAM. The idea is that nodes can be started up in batches. In our experimentation,
each node occupies, during startup, a maximum of 80\%/750=0.106\% of the available
RAM and $50\%/750=0.072\%$ after startup. Having, $80\% - 54\% = 26\%$ of space,  we could run a further batch of 243 nodes (totaling 993 nodes)
without occupying more than the $80\%$ of the RAM during startup and finally
occupying $(750+243)\cdot 0.072\%=71\%$ of the RAM totally available for containers.
This approach can be iterated by exploiting the remaining $80\% - 71\% = 9\%$ of space to run a further
batch of 84 nodes, and so on.

\section{Conclusions and Future Works}\label{sec:conclusions}

We showed a number of techniques targeted to solve several technical and methodological problems with the aim of enabling scalable and realistic emulation of blockchain networks.

We described our experience of adopting these techniques in two contexts: a
research experimentation and an emulation of a realistic production-like
network. While the machinery is quite complex, our experiences are promising. We
were able to run, on a single (virtual) machine, more than 3000 containers
executing a quite complex research-targeted blockchain and an Ethereum 2 network
consisting of 750 nodes, with the possibility grow above 1 thousand. Thanks to our time-inflation technique,
in our experiments, the 
RAM is the bottleneck (400GiB in our machine) and not the CPU. 

While we developed and tested our techniques for blockchain emulation, we
believe that most of the findings can be equally applied in other contexts, for
example for testing file sharing peer-to-peer networks or for large internet routing emulations.

There are a number of possible objectives for future works. 
\begin{enumerate}
\item A system that simplifies the setup could be very helpful for giving the
possibility to other research groups to reproduce  large scale emulations of
this kind.

\item Currently, our approach is limited to a single host. We intend to explore
the possibility of distributing containers among several machines adopting
technologies like Kubernetes~\cite{kubernetes}, possibly leveraging the Kathará
system~\cite{scazzariello2020kathara,scazzariello2021megalos}.

\item Usually blockchain experiments require the creation of a transaction
load and gathering data about the behavior of the network. 
A reusable tool supporting these tasks would be desirable.

\item Finally, blockchain experiments may need to setup a (possibly randomly generated)
overlay topology. A tool supporting this activity would also be desirable.

\end{enumerate}

\end{document}